\documentclass[11pt]{article}
\usepackage{blois,epsfig}

\def\be{\begin{equation}}
\def\ee{\end{equation}}
\def\bea{\begin{eqnarray}}
\def\eea{\end{eqnarray}}

\begin{document}
\title{Theories of Dark Energy with Screening Mechanisms}

\author{Justin Khoury}

\address{Center for Particle Cosmology, Department of Physics and Astronomy,\\ University of Pennsylvania,
  Philadelphia, PA , USA 19104}

\maketitle\abstracts{Despite the overwhelming evidence for the existence of dark energy and dark matter, 
their underlying fundamental physics remains unknown. This review article explores the tantalizing possibility that the dark sector
includes new light degrees of freedom that mediate long-range forces on cosmological scales. To ensure consistency with
laboratory and solar system tests of gravity, some screening mechanism is necessary to ``hide" these degrees of freedom locally.
I will focus on two broad classes of screening theories, chameleons and symmetrons, which rely respectively on the scalar field
acquiring a large mass or weak coupling in the presence of large ambient matter density.}

\section{Introduction}

The $\Lambda$-Cold Dark Matter ($\Lambda$CDM) standard model, which successfully accounts for a host of cosmological observations~\cite{bhuvrev}, will come under
increased scrutiny in this decade. The conjunction of the Large Hadron Collider, direct/indirect dark matter searches and forthcoming large-scale structure surveys will probe the nature of the dark sector with unprecedented accuracy from $10^{-16}$ to $10^{28}$~cm. While this is unlikely to result in a sweeping overthrow of our current picture, it may well reveal critical necessary addenda towards a complete fundamental description of cosmology. 

With this in mind, this article explores the tantalizing prospect that $\Lambda$CDM is only the simplest approximation to a dark sector endowed with a much
richer structure. Specifically, our main focus is the possibility that the dark sector includes new light degrees of freedom that couple to
both dark and baryonic matter with gravitational strength, thereby affecting the nature of gravity and the growth of structure on the largest scales. Naively the existence of light,
gravitationally-coupled scalars are ruled out by solar system tests of gravity. Over the last decade, however, it has been realized that scalar fields can in fact be clever
and evade detection from local experiments through {\it screening mechanisms}.

Aside from cosmology, these classes of theories find independent motivation in the vast experimental effort aimed at testing the fundamental nature of gravity at long wavelengths~\cite{will}. Viable screening theories make novel predictions for local gravitational experiments. The subtle nature of these signals have forced experimentalists to rethink the implications of their data and have inspired the design of novel experimental tests. The theories of interest thus offer a rich spectrum of testable predictions, from laboratory to extra-galactic scales. 

Only a handful of successful screening mechanisms have been proposed to date:

\begin{itemize}

\item {\it The Chameleon Mechanism}~\cite{cham1,cham1b,cham2,amol,cham3} operates whenever a scalar field couples to matter in such a way that its effective mass depends on the local matter density (Sec.~\ref{cham}). Deep in space, where the mass density is low, the scalar is light and mediates a fifth force of gravitational strength, but near the Earth, where experiments are performed, and where the local density is high, it acquires a large mass, making its effects short range and hence unobservable. Chameleon theories have, in particular, been central in developing viable $f(R)$ gravity theories~\cite{fR,fRrealistic}.  

\item {\it The Vainshtein Mechanism}~\cite{vainshtein,ags,ddgv} relies on derivative couplings of a scalar field becoming large in the vicinity of massive sources. These non-linearities crank up the kinetic term of perturbations, thereby weakening their interactions with matter. This mechanism is essential to the viability of massive gravity~\cite{vainshtein,ags,ddgv,claudia}, degravitation theories~\cite{degrav1,degrav2,degrav3}, brane-induced gravity models~\cite{DGP,cascade,cascade2,cascade3,cascade4} and galileon theories~\cite{galileon,galileon2,deser,multi1,multi2,nathan,koyama}.

\item {\it The Symmetron Mechanism}~\cite{symmetron,symmetronearlier} relies on the vacuum expectation value (VEV) of a scalar field that depends on the local mass density, becoming large in regions of low mass density, and small in regions of high mass density (Sec.~\ref{symm}). The coupling of the scalar to matter is proportional to the VEV, so that the scalar couples with gravitational strength in regions of low density, but is decoupled and screened in regions of high density. 

\end{itemize}
Below we will review the first and third of these mechanisms.

\section{Chameleon Field Theories}
\label{cham}

Chameleon scalar fields mediate a fifth force of gravitational strength between matter particles, with a range that decreases with increasing ambient matter density,
thereby avoiding detection in regions of high density~\cite{cham1,cham1b,cham2,cham3}. This is achieved within the framework of a general scalar-tensor theory
with scalar potential $V(\phi)$ and matter coupling $A(\phi)$:
\be
S = \int {\rm d}^4x\sqrt{-g}\left(\frac{M_{\rm Pl}^2}{2} R - \frac{1}{2}(\partial\phi)^2 - V(\phi)\right) + S_{\rm matter}\left[g_{\mu\nu}A^2(\phi) \right] \,.
\label{Scham}
\ee
One can in fact allow different couplings to the various matter fields, thereby explicitly violating the Equivalence Principle. For the purpose of this article, 
we focus on the simplest case of a universal, conformal coupling. The equation of motion for $\phi$ that derives from this action is
\be
\nabla^2\phi = V_{,\phi} - A^3(\phi)A_{,\phi} \tilde{T}\,,
\label{phigen}
\ee
where $\tilde{T} = \tilde{g}_{\mu\nu} \tilde{T}^{\mu\nu}$ is the trace of the energy-momentum tensor defined with respect to the
Jordan-frame metric $\tilde{g}_{\mu\nu} = A^2(\phi) g_{\mu\nu}$. Since matter fields couple minimally to $\tilde{g}_{\mu\nu}$, this stress tensor is covariantly
conserved: $\tilde{\nabla}_\mu  \tilde{T}^{\mu\nu} = 0$. (Although we focus on scalar fields for concreteness, chameleonic vector fields, such as gauged $B-L$,
have also been proposed~\cite{nelsonchamvec}.)

To study the field profile on solar system and galactic scales, we can approximate the metric in~(\ref{phigen}) as flat space, ignore time derivatives, and focus on 
the case of a non-relativistic, pressureless source. In terms of an energy density $\rho = A^3(\phi)\tilde{\rho}$ conserved in the Einstein frame, we obtain
\be
\nabla^2\phi = V_{,\phi} + A_{,\phi}\rho\,.
\label{phistat}
\ee
Thus, because of its coupling to matter fields, the scalar field is affected by ambient matter density, and is governed by an effective potential
\be
V_{\rm eff}(\phi) = V(\phi) +A(\phi)\rho \,.
\label{Veffcham}
\ee
For suitably chosen $V(\phi)$ and $A(\phi)$, this effective potential can develop a minimum  at some finite field value $\phi_{\rm min}$ in the presence of background matter density, where the mass of the chameleon field is sufficiently large to evade local constraints:
\be
m_{\rm min}^2 = V_{,\phi\phi}(\phi_{\rm min}) + A_{,\phi\phi}(\phi_{\rm min})\rho\,.
\ee
Assuming $A(\phi)$ is monotonically increasing, for concreteness, the general conditions that $V$ must satisfy are~\cite{cham1,cham1b,cham3}: (i) $V_{,\phi} < 0$ over the relevant field range,
in order to balance the potential against the density term; (ii) since $V_{,\phi\phi}$ typically gives the dominant contribution to $m_{\rm min}$, stability requires $V_{,\phi\phi}> 0$; (iii) in order for $m_{\rm min}$ to increase with $\rho$, we demand that $V_{,\phi\phi\phi} < 0$.

A prototypical potential satisfying these conditions is the inverse power-law form, $V(\phi) = M^{4+n}/\phi^n$. 
For the coupling function, a generic form that makes contact with Brans-Dicke theories is $A(\phi) \approx 1 + \beta\phi/M_{\rm Pl}$,
where we have used the fact that $\phi\ll M_{\rm Pl}$ over the relevant field range. The parameter $\beta$ is implicitly assumed to be~${\cal O}(1)$, corresponding to gravitational strength coupling. 
(Remarkably, much larger couplings  $\beta\gg 1$ are allowed by current constraints~\cite{shawpaper}, but one must be concerned with adiabatic instabilities in this regime~\cite{markandrachel}.)
The effective potential in this case is therefore given by, up to an irrelevant constant,
\be
V_{\rm eff}(\phi) = \frac{M^{4+n}}{\phi^n} + \beta \frac{\phi}{M_{\rm Pl}} \rho\,.
\ee
For $\beta> 0$, this displays a minimum at $\phi_{\rm min} \sim \rho^{-1/(n+1)}$. It follows that the mass of small fluctuations around the minimum, $m^2_{\rm min}\sim \rho^{(n+2)/(n+1)}$, is an increasing function of the background density, as desired. 

The tightest constraint on the model comes from laboratory tests of the inverse square law, which set an upper limit of $\approx 50\;\mu$m
on the fifth-force range assuming gravitational strength coupling~\cite{adel}. Modeling the chameleon profile in the E$\ddot{{\rm o}}$t-Wash set-up, and taking into account
that torsion-balance measurements are performed in vacuum, this constraint translates to an upper bound on $M$~\cite{cham1,cham1b}
\begin{equation}
M\; \lower .75ex \hbox{$\sim$} \llap{\raise .27ex \hbox{$<$}}\; 10^{-3}\;\;{\rm eV}\,,
\label{Mlim}
\end{equation}
which, remarkably, coincides with the dark energy scale.
This also ensures consistency with all known constraints on deviations from General Relativity, including post-Newtonian tests in the solar system and binary pulsar observations~\cite{cham1,cham1b,cham2}. The bound~(\ref{Mlim}) in turn implies a range of $\sim {\rm Mpc}$ in the cosmos --- too heavy to act as quintessence, but light enough to impact the growth of structure~\cite{cham3,wayneNbody}. 

Potentials with {\it positive} powers of the field, $V(\phi) \sim \phi^{2\alpha}$ with $\alpha$ an integer $\geq 2$, are also good candidates for chameleon theories~\cite{cham2}.
When the simplest case of $V(\phi) = \lambda \phi^4$ was initially considered, it was showed that existing laboratory constraints at the time were satisfied for $\beta = 1$ and $\lambda = 1$,
and that chameleons led to various signatures for future laboratory tests of the inverse-square-law~\cite{cham2,amol}. Subsequent analysis by the E$\ddot{{\rm o}}$t-Wash group~\cite{eotwash} excluded a
significant part of the parameter space.

\subsection{Thin-Shell Screening}
\label{thinshellcham}

The density-dependent mass immediately results in a further decoupling effect outside sufficiently massive objects, due to the so-called {\it thin-shell} effect. Consider a spherical source of
radius $R$ and density $\rho_{\rm in}$ embedded in a homogeneous medium of density $\rho_{\rm out}$. The corresponding effective minima of the effective potential will be respectively denoted by
$\phi_{\rm min-in}$ and $\phi_{\rm min-out}$. For a sufficiently massive source, the scalar field is oblivious to the exterior matter and is therefore pinned near
$\phi_{\rm min-in}$ in the core of the object. Of course, $\phi$ must deviate substantially from $\phi_{\rm min-in}$ near the surface of the object since $\phi$ must eventually reach the
asymptotic value $\phi_{\rm min-out}$ far away. Thus the gradient in $\phi$ builds up only within a thin-shell of thickness $\Delta R$ below the surface. Explicit calculations show that
\be
\frac{\Delta R}{R} = \frac{1}{6\beta M_{\rm Pl}} \frac{\phi_{\rm min-out} - \phi_{\rm min-in}}{\Phi_{\rm N}}\,,
\label{thin}
\ee
where $\Phi_{\rm N}$ is the surface Newtonian potential. In other words, the shell thickness is determined by the difference in $\phi$ values relative to the difference
in gravitational potential between the surface and infinity.  

Since field gradients are essentially confined to the shell, the exterior profile is suppressed by a thin-shell factor:
\be
\phi_{\rm screened} \approx  -\frac{\beta}{4\pi M_{\rm Pl}} \frac{3\Delta R}{R} \frac{Me^{-m_{\rm min-out} r}}{r} + \phi_{\rm min-out}\,.
\ee
The suppression factor $\Delta R/R$ can alternatively be understood intuitively as follows. Deep inside the source, the contribution to the exterior profile from
infinitesimal volume elements are Yukawa-suppressed due to the large effective chameleon mass in the core. Only the contributions from within the
thin shell propagate nearly unsuppressed to an exterior probe.

Clearly the thin-shell screening breaks down for sufficiently small objects. Imagine shrinking the source keeping the density fixed. Eventually, the cost in gradient energy
required to maintain the field difference between $\phi_{\rm min-in}$ and $\phi_{\rm min-out}$ becomes too large, and the scalar field no longer reaches $\phi_{\rm min-in}$ in the core of
the object. In this limit the thin-shell screening goes away, and the exterior profile takes on its usual form
\be
\phi_{\rm unscreened} \approx  -\frac{\beta}{4\pi M_{\rm Pl}} \frac{Me^{-m_{\rm min-out} r}}{r} + \phi_{\rm min-out}\,.
\label{noscreen}
\ee
The criterion for thin-shell screening to be effective is for the right-hand side of~(\ref{thin}) to be $\ll 1$. The effective
coarse-grained description of chameleon theories, including careful considerations of the thin-shell effect and averaging, has been derived~\cite{shawpaper}. 

\subsection{Observational Signatures}
\label{chamobs}

Clearly, the more massive the source, the easier it is to satisfy the thin-shell condition~(\ref{thin}), as expected. But note that the criterion also depends on the density
contrast --- for a given source, a denser environment implies smaller $\phi_{\rm min-out}$, which makes the thin-shell condition easier to satisfy. 
In particular, test masses that are screened in the laboratory may be unscreened in space. This leads to striking predictions for future satellite tests of gravity, such as the planned MicroSCOPE, Galileo Galilei and STEP missions. If~(\ref{Scham}) is generalized to include different chameleon couplings for different matter fields, then tests of the Equivalence Principle in orbit
might observe violations with $\eta \gg 10^{-13}$, in blatant conflict with laboratory constraints.
Meanwhile, from~(\ref{noscreen}) the total force (gravitational + chameleon-mediated) between unscreened particles is a factor of $1 + 2\beta^2$ larger than pure gravity, which would appear as ${\cal O}(1)$ deviations from $G_{\rm N}$ measured on Earth.

But even when the chameleon couples universally to matter fields, as in~(\ref{Scham}), the thin shell effect leads to {\it macroscopic} violations of the Equivalence Principle because objects of different mass have different {\it effective} coupling to the scalar~\cite{lamnic}. Effective violations of the Equivalence Principle also result in a host of astrophysical signatures~\cite{lamnic}. For example, (unscreened) dwarf galaxies in large void regions consist of unscreened HI gas and screened stars. The gas therefore feels both chameleon and gravitational attraction, whereas the stars only feel gravity. This should result in a systematic ${\cal O}(1)$ mismatch in the rotational velocities of these two tracers,
and hence a corresponding mismatch in mass estimates. 

Other interesting signals arise if the chameleon couples to the electromagnetic field
\be
\int {\rm d}^4x \sqrt{-g}e^{\beta_{\gamma}\phi/M_{\rm Pl}} F_{\mu\nu}F^{\mu\nu}\,.
\ee
Such a coupling results in photon-chameleon oscillations in the presence of an external magnetic field. The GammeV experiment~\cite{gammev}
searches for the afterglow~\cite{afterglow} that would result from trapped chameleons converting back into photons and has
so far excluded the range $5\times 10^{11} < \beta_\gamma < 6.4 \times 10^{12}$. The second-generation experiment, GammeV-CHASE, is currently taking data and will further improve on this bound~\cite{gammevnext}. Similarly, the ADMX microwave cavity was used recently to search for chameleons~\cite{admx}. Photon-chameleon mixing can also be detected astrophysically, through induced polarization in the spectrum of astrophysical objects and enhanced scatter in the X/$\gamma$-ray luminosity relation of active galactic nuclei (AGN)~\cite{clare}. 

\section{Symmetron Fields}
\label{symm}

A second mechanism for hiding scalar fields was proposed recently in the context of symmetron field theories~\cite{symmetron,symmetronearlier}. 
Although the symmetron technology and building blocks are similar to those of chameleon theories, the physics of the screening mechanism and
its phenomenological consequences are dramatically different. In particular, unlike in chameleon theories, the symmetron has a small mass everywhere
and therefore mediates a long-range force~\cite{symmetron}.

In the symmetron mechanism, the vacuum expectation value (VEV) of the scalar depends on the local mass density, becoming large in regions of low mass density, and small in regions of high mass density. In addition, the coupling of the scalar to matter is proportional to the VEV, so that the scalar couples with gravitational strength in regions
of low density, but is decoupled and screened in regions of high density. 

The starting point is the general scalar-tensor theory~(\ref{Scham}). The screening mechanism is achieved through the interplay of a symmetry-breaking potential, 
\be
V(\phi)=-{1\over 2}\mu^2\phi^2+{1\over 4}\lambda\phi^4\,,
\label{Vsymm}
\ee
and $Z_2$-invariant universal conformal coupling to matter
\be
A(\phi) = 1+{\phi^2\over 2M^2} + {\cal O}(\phi^4/M^4)\,.
\label{Asymm}
\ee
The field range of interest satisfies $\phi \ll M$, as we will see shortly, thus higher-order terms in $A(\phi)$ are negligible. 
For non-relativistic matter, $T^\mu_{\;\mu}\approx -\rho$, the effective potential is thus given by, up to an irrelevant constant,
\begin{equation}
V_{\rm eff}(\phi) = {1\over 2}\left({\rho\over M^2}-\mu^2\right)\phi^2+{1\over 4}\lambda\phi^4\,.
\label{Veff}
\end{equation}
Therefore, whether the $Z_2$ symmetry is spontaneously broken or not depends on the local matter density.
In vacuum ($\rho = 0$), the scalar field acquires a VEV, $\phi_0 \equiv \pm \mu/\sqrt{\lambda}$, thereby breaking the symmetry spontaneously.
In the presence of sufficiently high matter density, such that $\rho > M^2\mu^2$, the effective potential is instead minimized at $\phi = 0$, where the symmetry is restored. 

An essential feature is that the coupling of the symmetron to matter is $\sim\phi^2\rho /M^2$.  Perturbations $\delta\phi$ around a local background value $\bar{\phi}$, as probed by experiments, therefore couple as
\begin{equation}
{\bar{\phi} \over M^2}\delta\phi \ \rho\,.
\label{coupling}
\end{equation}
That is, the coupling, which determines the strength of the symmetron-mediated force, is proportional to the local background $\bar{\phi}$.  In high-density, symmetry-restoring environments, such as our galaxy, we have $\bar{\phi} \approx 0$, and the symmetron force is feeble. In rarified, symmetry-breaking environments, such as the cosmos, the
symmetron force can be of gravitational strength. 

The symmetron naturally takes the form of an effective field theory. The potential comprises the most general renormalizable terms invariant under
the $Z_2$ symmetry $\phi\rightarrow -\phi$. The coupling to matter is the leading such coupling compatible with the symmetry.  It is non-renormalizable, suppressed by the mass scale $M$, thus the symmetron is an effective theory with cutoff $M$.  In fact, this cutoff is near the GUT scale, so any GUT theory with a low energy scalar might be expected to yield a symmetron-type Lagrangian at low energies. From this point of view, symmetron theories are more natural-looking than chameleon models. As with chameleons, however, the coupling to matter generates large quantum corrections to the potential which must be fine-tuned away. 

The case of interest is when the field becomes tachyonic around the current cosmic density~\cite{symmetron}: $H_0^2M_{\rm Pl}^2 \sim \mu^2M^2$. This fixes $\mu$ in terms of $M$,
and hence the mass $m_0$ of small fluctuations around the symmetry-breaking vacuum:
\be
m_0= \sqrt{2} \mu \sim \frac{M_{\rm Pl}}{M} H_0\,.
\label{muvalue}
\ee
Local tests of gravity, as we will see, require $M \;  \lower .75ex \hbox{$\sim$} \llap{\raise .27ex \hbox{$<$}}  \;10^{-3}M_{\rm Pl}$. Hence the range $m_0^{-1}$ of the symmetron-mediated force
in vacuum is $ \lower .75ex \hbox{$\sim$} \llap{\raise .27ex \hbox{$<$}}$~Mpc. Meanwhile, if this extra force is to be comparable in strength to the gravitational force, then from~(\ref{coupling}) we must impose
$\phi_0/M^2 \sim 1/M_{\rm Pl}$, that is,
\be
\phi_0\equiv \frac{\mu}{\sqrt{\lambda}} \sim \frac{M^2}{M_{\rm Pl}}\,.
\label{vev}
\ee
Together with~(\ref{muvalue}), this implies $\lambda \sim M_{\rm Pl}^4H_0^2/M^6 \ll 1$.
(For $M=10^{-3}M_{\rm Pl}$, in particular, this gives $\lambda = 10^{-102}$ --- the symmetron is extremely weakly coupled.)
We see immediately from~(\ref{vev}) that $\phi_0 \ll M$, hence the field range of interest lies within the regime of the effective field theory,
and ${\cal O}(\phi^4/M^4)$ corrections in~(\ref{Asymm}) can be consistently neglected, as claimed earlier.

\subsection{Symmetron Thin-Shell Screening}

Symmetron solutions around a source display a thin-shell effect closely analogous to the chameleon behavior discussed in Sec.~\ref{thinshellcham}. Consider once again the ideal case of a static, spherically-symmetric source of homogeneous density $\rho >  \mu^2M^2$. For simplicity, we assume that the object lies in vacuum, so that the symmetron tends to its symmetry-breaking VEV far away: $\phi\rightarrow\phi_0$ as $r\rightarrow \infty$. 

For a sufficiently massive source, in a sense that will be made precise shortly, the solution has the following qualitative behavior. Deep in the core of the object, 
the symmetron is weakly coupled to matter, since the matter density forces $\phi\approx 0$ there. Near the surface, meanwhile, the field must grow
away from $\phi = 0$ in order to asymptote to the symmetry-breaking VEV far away. The symmetron is thus weakly coupled to the core of the object,
and its exterior profile is dominated by the surface contribution. In other words, analogously to chameleon models, there is a thin shell screening effect
suppressing the symmetron force on an external probe.

Explicit calculations show that whether screening occurs or not depends on the parameter~\cite{symmetron}
\be
\alpha \equiv \frac{\rho R^2}{M^2} = 6\frac{M_{\rm Pl}^2}{M^2}\Phi_{\rm N} \,.
\label{alp}
\ee
Objects with $\alpha\gg 1$ display thin-shell screening, and the resulting symmetron-mediated force on a test particle
is suppressed by $1/\alpha$ compared to the gravitational force. Objects with $\alpha \ll 1$, on the other hand, do not have a thin shell --- the symmetron gives an ${\cal O}(1)$ correction to the gravitational attraction in this case. 

\subsection{Tests of Gravity and Observational Signatures}

Since the symmetron-mediated force is long-range in all situations of interest, and because the symmetron couples to matter universally,
the relevant tests of gravity are the same that constrain Brans-Dicke theories: solar system and binary pulsar observations. It turns out that a necessary condition
is for the Milky Way galaxy to be screened~\cite{symmetron}: $\alpha_{\rm G} \;\lower .75ex \hbox{$\sim$} \llap{\raise .27ex \hbox{$>$}} \; 1$. Since $\Phi \sim 10^{-6}$ for the galaxy, in this case~(\ref{alp}) implies
\be
M \;  \lower .75ex \hbox{$\sim$} \llap{\raise .27ex \hbox{$<$}}\;  10^{-3}M_{\rm Pl}\,.
\ee
It turns out that this condition is sufficient to satisfy all current constraints. Indeed, with
$M = 10^{-3}M_{\rm Pl}$, the symmetron predictions for time-delay, light-deflection, perihelion precession of Mercury and Nordvedt effect
are all comparable to current sensitivity levels and therefore detectable by next-generation experiments~\cite{symmetron}. Note that pushing $M$ to larger
values is also desirable cosmologically, since the range of the symmetron force grows with $M$.
In particular, from~(\ref{muvalue}), $\mu^{-1}\;  \lower .75ex \hbox{$\sim$} \llap{\raise .27ex \hbox{$<$}} \; {\rm Mpc}$ for $M \;  \lower .75ex \hbox{$\sim$} \llap{\raise .27ex \hbox{$<$}} \;10^{-3}M_{\rm Pl}$.

The symmetron is observationally distinguishable from other screening mechanisms. In chameleon theory, as discussed in Sec.~\ref{cham},
the tightest constraints come from laboratory tests of the inverse square law. Once these are satisifed, however, the predicted solar system deviations
turn out to be unobservably small. In contrast, as mentioned above the symmetron predictions for solar system tests are just below current constraints.
On the other hand, chameleon and symmetron models have in common the prediction of macroscopic violations of the Equivalence Principle, which
can show up in various astrophysical observations~\cite{lamnic}. We note in passing that in the case of Vainshtein screening, brane-induced gravity and degravitation models
predict modifications to the Moon's orbit that are within reach of next generation Lunar Laser Ranging observations~\cite{moon,niayeshghazal}, but light-deflection
and time-delay signals are negligible. 

\section{Conclusions}

The standard $\Lambda$CDM model of cosmology will be tested with unprecedented accuracy over the coming decade.
In anticipation of potential surprises that may be lurking at the Large Hadron Collider or awaiting cosmological surveys
such as the Dark Energy Survey and the Large Synoptic Space Telescope, it is prudent to explore a broader scope of microphysics and associated phenomena beyond the standard paradigm.
This article has surveyed recent theoretical developments that open the exciting possibility of dark energy interacting with
both dark and baryonic matter, thereby mediating additional long range forces on cosmological scales. 
These ideas rely on screening mechanisms to ensure consistency with local tests of gravity. We have reviewed
two broad classes of theories that exhibit screening --- chameleons and symmetrons --- and result in a host of
striking experimental signatures, from laboratory to cosmological scales~\cite{wyman}.


\section*{Acknowledgments}
This work is supported in part by funds from the University of Pennsylvania and by an Alfred P. Sloan Research Fellowship.

\end{document}